\documentclass[pre,a4paper,superscriptaddress,twocolumn,showpacs,amsmath,amssymb,floatfix]{revtex4}

\usepackage{graphicx}
\usepackage{amssymb}
\usepackage{float}
\usepackage{hyperref}

\usepackage{color}

\def\eps{\varepsilon}

\begin{document}
	
\title{Thermodynamic statistics of given names in USA and France}
	
\author{Klaus M. Frahm}
\affiliation{{\mbox Univ Toulouse, CNRS, Laboratoire de Physique Th\'eorique, 
Toulouse, France}}
\author{Dima L. Shepelyansky}
\affiliation{{\mbox Univ Toulouse, CNRS, Laboratoire de Physique Th\'eorique, 
Toulouse, France}}
\affiliation{\mbox{{\bf Author to whom correspondence should be addressed: dima@irsamc.ups-tlse.fr}}}

\date{August 6, 2026}

\begin{abstract}
  Using official government data sets of USA and France we
  analyze the occurrence/frequency/popularity distributions of given names
  on a time scale of more than 100 years.
  These distributions are characterized through the Lorenz and Pareto curves
  broadly used in the analysis of wealth inequality in the world.
  These curves remain stable during the considered time period 
  with the Gini coefficient remaining in the narrow range 0.85-0.95.
  As for the case of wealth inequality, we show that
  the distributions of names are well described by the
  Rayleigh-Jeans (RJ) thermalization and condensation phenomenon
  well studied in various physical systems. The RJ thermalization results
  from two integrals of motion being analogous to energy and probability norm
  conservation in physical systems with energy states corresponding
  to popularity levels of names. Time correlations between names
  are also determined showing their stability until the middle of the 
  twentieth century and a significant change after that.
\end{abstract}

%

\maketitle

{\bf Exploiting  government data bases, 
the statistical properties of given names in USA and France are analyzed 
for a period of over hundred years. 
The Lorenz and Pareto distributions are shown to be
very stable on this time scale implying
a steady-state regime of the system.
We show that this regime is well described by 
Rayleigh-Jeans (RJ) thermalization and condensation
reproducing accurately the Lorenz and Pareto curves 
from these data bases.
We argue that the thermalization appears due to effective
nonlinear interactions that influence 
the name choice process. 
The distribution of names has high values of Gini coefficient
with a huge phase of names condensation at low name popularity
and a tiny fraction of names being highly popular.
We trace an analogy with the
wealth inequality distribution in the world countries
also described in the framework of RJ condensation.
The statistical correlations
for a group of most popular names change 
significantly at the middle of the twentieth century.
}

\section{Introduction} 
\label{sec1}

\begin{figure}[h]
\begin{center}
\includegraphics[width=0.95\columnwidth]{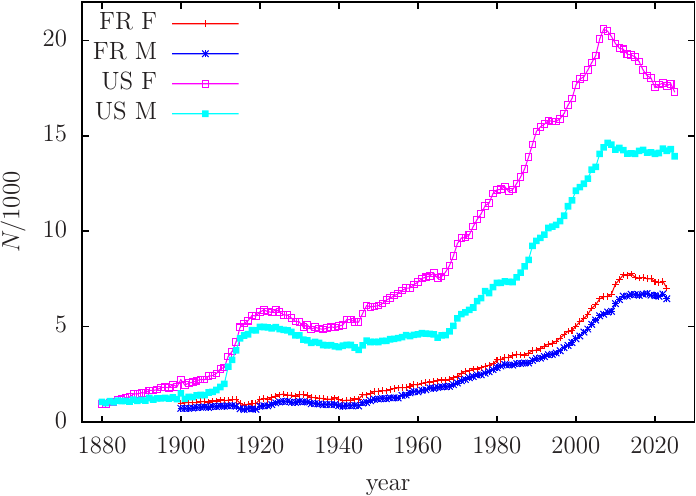}%
\end{center}
\caption{\label{fig1}
Year dependence of the number of given names in the data sets of \cite{govus,govfr} 
for the four cases of FR and US, both for female (F) and male (M) names. 
See also Appendix Fig.~\ref{figA1}, where normalized values of this number with respect 
to the year 1960 are shown. }
\end{figure}

There is a high public interest to the distribution and time evolution 
of the most popular baby names of a given year 
both in USA and France. Such names are regularly reported by public media
each years as e.g. for USA \cite{mediaus}
and for France (FR) \cite{mediafr} in the year 2025. 
Usually media highlight top ten or
a few tens of names for girls and boys.
However, the statistical government organizations
in USA and FR accurately register baby names
and make them available for public.
In particular, the frequencies (occurrences/popularity $f_m$) of given 
new born names $m$ in USA 
for each year from 1880 up to 2025 are freely
accessible at \cite{govus}
and similarly for France from 1900 up to 2023 at \cite{govfr}.
The dependence of the total number $N$ of Female (F) and Male (M) names
on years for USA and FR are shown in Fig.~\ref{fig1}.
To safeguard privacy, the US data \cite{govus} restricts 
the names to those with at least 5 occurrences,
for FR names \cite{govfr} it is at least 3 occurrences. 

The statistical analysis of the distribution of names,
linked to different cultures, over a country territory
also attracts a significant public interest as it is e.g. the case
for results presented in \cite{fourquet}.

In this work, we analyze the global statistical
properties of distributions of frequencies of given names
constructed from US and FR data sets \cite{govus,govfr} on a time scale
of 145 or 123 years. To characterize these
distributions we follow the approach broadly used in studies
of wealth inequality in different countries
and in the whole world \cite{piketty1,piketty2,boston}. 
In this approach the distribution of wealth
of households is described by the Lorenz curve 
\cite{lorenz,boston} which gives the dependence of cumulated 
normalized wealth $0\leq w \leq 1$ on the cumulated normalized fraction of
population or households $0 \leq h \leq 1$.
The equipartition of wealth corresponds to
the diagonal $w=h$ and the doubled area between diagonal
and the Lorenz curve $w(h)$ determines the Gini coefficient
$0 \leq G \leq 1$  \cite{gini,boston}.
For the whole world in 2021 the high value $G=0.889$ \cite{wikigini} 
indicates a huge wealth inequality in the world, i.e. 
50\% of the population owns only 2\% of total wealth while 
10\% (1\%) of population owns 75\% (38\%) of total wealth
\cite{piketty2}.

We use the concepts of this wealth inequality approach for the 
statistical analysis of first name distributions in USA and FR
considering that a name frequency $f_m$ is analogous to wealth
$w_m=f_m$ of household $m$ 
and the number of names (name index) is analogous to
a household index $m$. Our analysis shows that the Lorenz distributions
of names are also characterized by a huge inequality
when approximately  50\% of all names have only 1.2\% of cumulated 
frequencies of all names and top 10\% of all names
capture 85\% of total name frequencies (e.g. for US female names in 2025).
We also analyze the cumulative distribution function of names, known as
{\rm Pareto} distribution \cite{pareto} (see below), 
that highlights in a better way the distribution
at high frequencies.

As for the case of wealth inequality, we argue that the
physical grounds of this phenomenon
are related to the  Rayleigh-Jeans (RJ) thermalization and condensation
at low energy states for
interacting classical fields/households/agents/players \cite{wth1,wth2}. 
According to the Wealth Thermalization Hypothesis (WTH)
introduced in \cite{wth1} wealth is analogous to energy of system states
(or states of social stratification in a society \cite{wth3})
and, as in physics, its distribution has the RJ form.
The thermalization process is characterized by two conserved integrals
of motion being the total energy/wealth (E) 
and the total probability norm ($\eta$).  
The important role of these two integrals of motion has been discussed in 
\cite{boghosian1,boghosian2}.
In presence of these two conserved quantities the RJ thermal
distribution is characterized by system temperature $T$
and chemical potential $\mu$ related to
energy total $E$ and norm $\eta$ \cite{landau,zakharovbook}.

At low values of total system energy (or temperature)
the RJ thermalization is characterized by a condensation
of huge fraction of total norm
on low energy states. This phenomenon has also been studied
and experimentally observed in multimode optical fibers
\cite{wabnitz,picozzi1,chrisrep,picozzi2,ourfiber}.
The RJ condensation is a special case of a more 
generic phenomenon known in statistical mechanics as
constraint-driven condensation \cite{trizac,satya,marsili}
that is universal and
exists for such systems as
coalescence in granular media, jamming in traffic, 
gelation in networks \cite{satya}
and financial data analysis \cite{marsili}.

Recently, it was shown that
the RJ condensation also describes the distributions
of energy consumption and carbon emission over world countries \cite{energy},
and the voting distributions for EU elections \cite{election}.
On the basis of the findings of \cite{wth1,wth2,wth3,energy,election}
and results presented in this work
we argue that the RJ thermalization and condensation
can also be applied to the frequency distributions of given names in US and FR.
The effective interactions between names appear via discussions between people,
via books, magazines, radio and other types of media.
The specific mechanism of these interactions
is not important for the issue of thermalization. 
This is for example similar to 
atoms in a 3D box where almost any interaction 
leads to the thermal Boltzmann-Gibbs distribution
of kinetic energies 
with the average value $3T/2$ for each atom
(with the Boltzmann constant taken as unity) \cite{landau}.

The article is constructed as follows: In 
Section \ref{sec2}, we recall some basic features of 
the RJ thermalization theory and the related construction 
of Lorenz and Pareto curves in this context. Section \ref{sec3}, 
presents results for the two considered data sets \cite{govus,govfr} 
and Section \ref{sec4} shows that the real data can be very well described 
by RJ thermalization with a specific model spectrum. Section \ref{sec5} 
provides some insights about time correlations of certain individual 
names or groups of names and the discussion of the results is given in 
\ref{sec6}. Additional material and figures are presented in the Appendix 
(Section \ref{secapp}).

\section{RJ thermalization and condensation}
\label{sec2}

For the classical evolution of fields or particles with two integrals 
of motion being energy and probability norm (or number of particles)
the Rayleigh-Jeans (RJ) thermal distribution over linear eigenmodes of
system Hamiltonian $H$ is given by \cite{landau,zakharovbook}: 
\begin{equation}
\rho_m = \frac{T}{E_m-\mu} \; ({\rm RJ}) 
\label{eqrj}
\end{equation}
where $\rho_m \ge 0$ is the average occupation number of a 
mode at energy $E_m$. Here $T$ and $\mu$ are the temperature and chemical 
potential. Their values are determined by the
conservation relations for energy
$E=\sum_m E_m \rho_m $ and norm
$\eta = \sum_m \rho_m =1$ where $\rho_m$ values are given by (\ref{eqrj}).
It is assumed that the nonlinear interactions between classical modes 
are weak and do not affect these relations. For a given spectrum $E_m$
the average system energy is  $E\in[E_0,E_{N-1}]$
where $N$ is the total number of modes (or names).
The values of $T$ and $\mu$ can be computed numerically
with relation (\ref{eqrj}) which implies $T=(E-\mu)/N$
(see e.g. \cite{wth1,wth2,rmtprl} for specific technical details on this).
Here, we consider the regime of positive temperatures $T>0$ when 
$\mu<E_0$ even though negative $T$ values are also possible with $\mu>E_{N-1}$ 
for sufficiently large values of $E$ (see \cite{wth1,rmtprl}).
We note that the RJ thermal distribution can also be obtained 
as a limiting case of the quantum Bose-Einstein distribution
for boson particles in a high temperature  regime when
$(E_m -\mu) \ll T$ \cite{landau}.

\begin{figure}[h]
\begin{center}
\includegraphics[width=0.95\columnwidth]{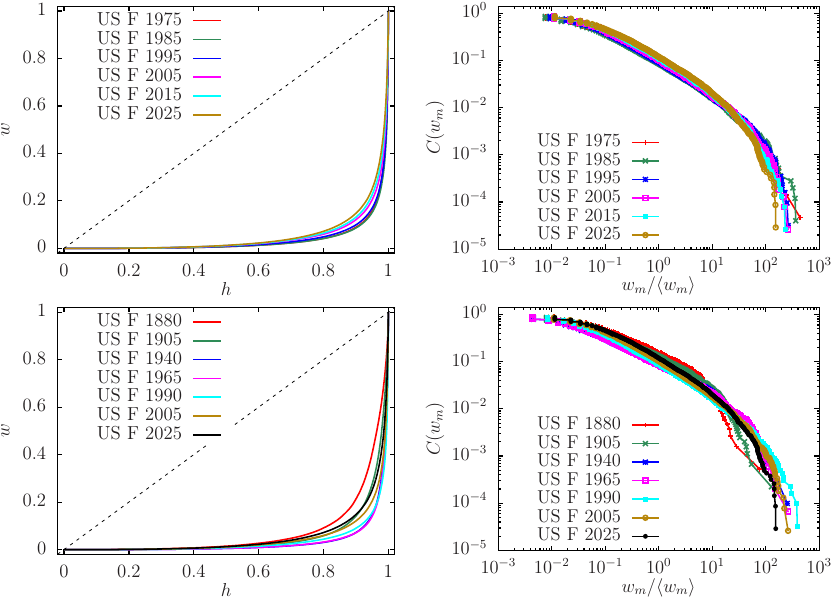}%
\end{center}
\caption{\label{fig2}
{\em Left:} US female name Lorenz curves 
for the 6 years 1975, 1985, 1995, 2005, 2015, 
2025 (top) and the 7 years 1880, 1905, 1940, 1965, 1990, 2005, 2025 (bottom). 
The $x$-axis corresponds to the 
cumulated fraction of households/names ($h$) and the $y$-axis to
the cumulated fraction of wealth/name frequency ($w$). 
The dashed black line 
corresponds to the line of perfect equipartition $w=h$. The Gini coefficients 
for the years 1975 to 2025 (1880 to 2025) are 
$G=0.93,0.934,0.925,0.911,0.903,0.894$ 
($G=0.854,0.895,0.938,0.939,0.931,0.911,0.894$).
{\em Right:} Pareto curves $C(w_m)$ for the same data where 
$C(w_m)$ represents the fraction of names with a name frequency 
larger than $w_m$ (analogous to wealth). 
The $x$-axis corresponds to rescaled 
values $w_m/\langle w_m\rangle$ where $\langle w_m\rangle = w_s$ is the 
average name frequency which takes for the years 
1975 to 2025 (1880 to 2025) the values 
$\langle w_m\rangle = 132,133,107,91.2,88.3,87.9$ 
($\langle w_m\rangle = 91.6,126,222,229,120,91.2,87.9$). 
See also Appendix Fig.~\ref{figA2} for the case of US male names. }
\end{figure}

For small energies $E\sim E_1-E_0$ and  temperatures values $T$ 
a macroscopic part of the total probability norm $\eta$ 
is condensed at the lowest energy modes $E_k$.
Following the WTH \cite{wth1}, we argue that
this condensation is at the origin of huge wealth inequality
in the world. In this approach
it is assumed that the wealth of individual households
$w_m$ is analogous to the system mode energies $E_m=w_m$.
Here we extend this analogy assuming that the name frequencies
$f_m=w_m=E_m$. More features of RJ condensation are described
in \cite{wth1,wth2,wth3,ourfiber,rmtprl}.

From the RJ  probabilities $\rho_m$ (\ref{eqrj})
we construct the Lorenz curve 
by computing the cumulated normalized 
fraction of households as $h(m) =\sum^{m-1}_{k=0} \rho_k$
and the cumulated normalized wealth fraction 
$w(m) = \sum^{m-1}_{k=0} w_k \rho_k/w_s$
where $w_k=E_k=f_k$ is the individual wealth/energy
(or name frequency) associated to a given agent $k$.
Here $w_s=f_s$ is the average wealth given by 
$w_s = \sum w_m\rho_m$  corresponding to the energy parameter $E$ in the 
RJ approach. This procedure gives the Lorenz curve as $N+1$ data points 
$(h(m),w(m))$ for $0\le m\le N$ such that $h(0)=w(0)=0$ and $h(N)=w(N)=1$ 
and it is invariant with respect to a simple energy 
rescaling $E_m\to\alpha E_m$, $E\to \alpha E$. Here the relevant 
parameter is the rescaled energy $\varepsilon = w_s/B=E/B$ where 
$B=E_{N-1}$ is the energy band width of the spectrum (in this work 
we also assume $E_0=0$). 

To compare a real Lorenz curve with the RJ theory (\ref{eqrj})
we suppose that names or household are represented by  a certain number
of thermalized agents $N$ distributed over
energy levels $E_k$ being in the energy band $B$.
The simplest assumption is that the density of energy states
$\nu(k) =d k/d E_k$ is constant so that
$0 \leq E_k=k/N < B=1$ for $k=0,1,2,\ldots,(N-1)$.
Such a case corresponds to the RJ standard (RJS) model
studied in \cite{wth1,wth2,ourfiber}. For small $\eps \ll 1$ values
the Lorenz curves obtained from this model agree qualitatively 
with real Lorenz curves for various systems. However, a more refined 
agreement is obtained in the framework of the RJ 
extended (RJE) model where the density of states decreases at high revenues:
$\nu(E_k)=dk/dE_k = N(e^a-1)/[a(1+(e^a-1)E_k]$, 
with $E_k=(e^{ak/N}-1)/(e^a-1)$ 
and the energy band width is $B = E_{N-1}\approx 1$.
This model depends on an additional parameter $a$
and in the limit $a \to 0$ it is reduced to the RJS model.
To reproduce the Lorenz curve with the RJ theory (\ref{eqrj}), 
we apply the above procedure with a sufficiently large number of agents
$N$ (usually $N=10^4$ or even more).  Then multiple agents describe
real values $\rho_m, w_m$ for a given household or name frequency $f_m=w_m$.
To compare RJ results with real data 
we fix the parameter $\eps$
by the condition that the RJ Lorenz curve has the same Gini coefficient 
as the real data Lorenz curve. 
For the RJE model, the parameter $a$ is fixed by the condition
that the distance between the RJE and the real Lorenz curve is minimal.

\begin{figure}[h]
\begin{center}
\includegraphics[width=0.95\columnwidth]{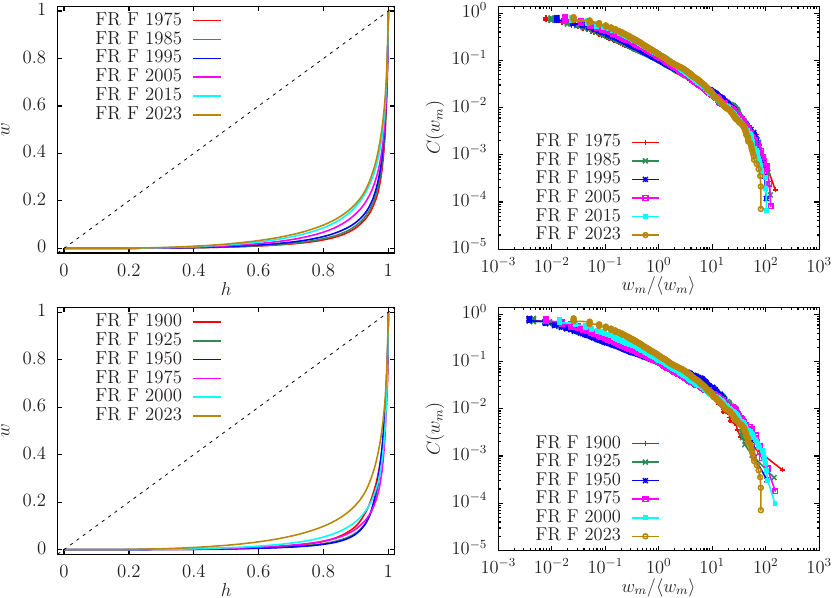}%
\end{center}
\caption{\label{fig3}
FR female name Lorenz and Pareto curves for the 6 
years 1975, 1985, 1995, 2005, 2015, 2023 (top) and the 6 years 
1990, 1925, 1950, 1975, 2000, 2023 (bottom) in 
the same style as Fig.~\ref{fig2}. The Gini coefficients 
for the years 1975 to 2023 (1900 to 2023) are 
$G=0.931,0.925,0.916,0.898,0.877,0.866$ 
($G=0.923,0.934,0.934,0.931,0.914,0.866$)
and the values of $\langle w_m\rangle = w_s$ are 
$\langle w_m\rangle=130,103,80.5,56.9,43.8,39.1$ 
($\langle w_m\rangle=237,221,265,130,71.6,39.1$).
See also Appendix Fig.~\ref{figA3} for the case of FR male names. }
\end{figure}

Another description of wealth inequality in addition to the Lorenz curve
can be obtained by the so-called Pareto distribution \cite{pareto}
that highlights properties of high revenues in a better way.
It is given as the cumulative distribution function (CDF) 
$C(w_m)$ which determines the fraction of households (names)
having a wealth (frequency) larger than $w_m=f_m$. This quantity is 
broadly used in economy (see e.g. \cite{yakovenko1,yakovenko2}) 
and it can be directly obtained 
by drawing $C(w_m)=1-h(m)$ versus $w_m$.
Usually it is expected that the Pareto distribution
has an algebraic decay. Here we show that it is not necessary
the case but we still use notation
{\it Pareto distribution or curve} in this work.
The dependence of Pareto curve is scale dependent
and for comparison with real data we
use the ratio $w_m/\langle w_m\rangle$ where
$\langle w_m\rangle$ is the average wealth or frequency value.
For the  RJS/RJE models we have $\langle w_m\rangle=\eps$ since $B=1$ and 
$E_m=w_m=f_m$. 

We mention, that while RJS and RJE Lorenz and Pareto curves can be 
efficiently computed using the above construction procedure for spectra 
with finite $N$ (e.g. $N=10^4$), there are also analytic explicit expressions 
of these curves for the continuous limit $N\to\infty$ which agree very 
well with those for finite but large $N$ (see \cite{wth2} for details). 
In this work, shown RJE curves were all computed from these analytic 
expressions. 

We note that in early works on thermodynamic description of
money, wealth or income distribution
it was argued that this distribution is given by the Boltzmann-Gibbs 
(BG) thermal distribution \cite{yakovenko1,yakovenko2}. 
In \cite{wth2,energy}, 
we gave arguments showing that it is not the case
and that the RJ thermalization (\ref{eqrj})
gives the right description
of real Lorenz and Pareto curves.

\begin{figure}[h]
\begin{center}
\includegraphics[width=0.95\columnwidth]{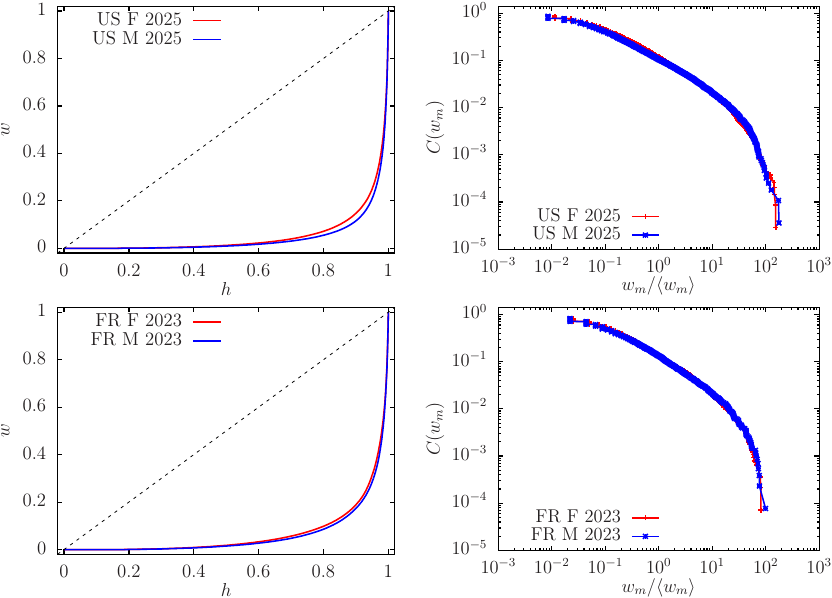}%
\end{center}
\caption{\label{fig4}
Comparison of Lorenz curves (left panels) and Pareto curves (right panels) 
between female names (red curves) and male names (blue curves). 
Top (bottom) panels correspond to data of US 2025 (FR 2023). 
See also Appendix Fig.~\ref{figA4} for the cases of US 1880 and FR 1990. }
\end{figure}

\section{Data sets, Lorenz and Pareto curves} 
\label{sec3}

The data sets for given names of new born and their
frequencies (occurrences) $f_m$
are available for 146 years (1880 to 2025)for US \cite{govus}
and 124 years (1900 to 2023) for FR \cite{govfr}. In both cases separate data 
for female and male names are provided. 

The dependence of the number of female and male names $N$ on years
is shown in Fig.~\ref{fig1} for both US and FR. 
Except for the period after 2005 (and 1910\,-1930 for US), the 
time dependence can be approximately fitted by an exponential 
growth $N(t)\sim e^{\lambda t}$ where $\lambda\approx 0.02$ 
and $t=$ year value corresponding to a typical yearly growth 
of $\approx 2$\% (see Appendix for more precise fit values of $\lambda$). 

For US the number of (both female/male) names $N$ 
are by a factor $R \approx 5.3 (1920); 2.8 (2005); 2.4 (2023)$
larger then for FR . These values are visibly higher (for 1920) 
or lower (for 2005 and 2023) than the population ratio
being $R_p \approx 2.7 (1920); 4.8 (2005); 4.9 (2023)$
(data from US Census Bureau and INSEE FR).
Also around the years 1920\,-1924 the number of names
in US has a maximum while for FR it has a minimum for 
1914\,-1918. 
We attribute this to the effect of World War I
that enhanced immigration to the US 
and reduced the birth rate in FR. 

More generally, the high cultural diversity due to 
the historically high immigration, also explains 
the larger numbers of used names for new born in the US. 
We also note that the four curves of Fig.~\ref{fig1} collapse 
on one curve, especially for the period of 1945\,-2005, if they 
are rescaled by the respective value of 1960 (see Appendix Fig.~\ref{figA1}) 
which is also coherent with the similar growth rate of $\approx 2$\% per 
year for this period. 

In this work, our main goal is to analyze the data of \cite{govus,govfr} with 
respect to a possible RJ thermalization, and in particular if the 
corresponding Lorenz and Pareto curves constructed from this data, 
as described in Section \ref{sec2}, can be matched to the RJE model. 
For this, we choose as usually the spectral ground state energy $E_0=0$, 
which means that the energy/name frequencies values $E_m=w_m$ are shifted 
according $w_m=f_m-f_{\min}$ where $f_{\min}$ is the minimal value 
which is either $5$ for the US or $3$ for the FR data. In this way, 
the minimal (shifted) frequency values are $w_0=0$ in coherence the 
model spectra of the RJS/RJE models and this corresponds essentially 
to a shift in the value of chemical potential $\mu$. 
In the following, we use these 
shifted frequency values, to construct Lorenz and Pareto curves. 

The maximal (shifted) frequencies $w_m$ of US names are between 
$7060/9650$ for 1800 and $13539/20813$ in 2025 for F/M.
For FR, we have the values $48710/14094$ for 1900 and $3174/4524$ in 2023
for F/M. These rather high values ensure a 
good statistical accuracy for Lorenz and Pareto curves.

In Fig.~\ref{fig2}, we compare the Lorenz and Pareto curves for US female 
names for two selected sets of years, being either 1975 to 2025 (step of 10 
years) or for the full period between 1880 to 2025 (steps of 25 or 35 years). 
The Lorenz curves for the shorter time interval are indeed very close with 
very similar Gini coefficients in the interval $G\in[0.894,0.934]$ while 
for the full time interval the deviations are a bit larger and with 
values of $G\in[0.854,0.939]$. 
The average frequency values $\langle w_m\rangle$, used for the 
computation of the Pareto curves (shown in a double logarithmic 
representation) are typically $\sim 10^2$ with minimal (maximal) value 
of 87.9 (for 2025) and 228 (for 1965). The largest values of the 
ratio $w_m/\langle w_m\rangle$ are typically between $10^2$ and 
$5\times 10^2$ with probabilities $C(w_m)\sim 1/N$ slightly above $10^{-5}$ 
which is coherent with the maximal frequency values given above and 
also the $N$ values of Fig.~\ref{fig1}. 

Fig.~\ref{fig3}, shows similar data for FR female names 
for the two sets 1975 to 2023 (steps of 10 or 8 years) 
and 1990 to 2023 (steps of 25 or 23 years). The stability of 
Lorenz and Pareto for the different periods is comparable to the US case. 
The values of the Gini coefficients are between 0.866 (2023) and 
0.934 (1925, 1950) while $\langle w_m\rangle$ is typically 
between 39.1 (2023) and 265 (1950). The maximal ratio 
$w_m/\langle w_m\rangle$ is typically $\sim 10^2$ with 
probabilities $C(w_m)\sim 1/N\sim 10^{-4}$. 

Appendix Figs.~\ref{figA2} and \ref{figA3} show similar data for 
the case of male US (Fig.~\ref{figA2}) and male FR (Fig.~\ref{figA3}) names. 

Globally, the Lorenz curves are quite stable for the shorter period 
1975 to 2025/2023 and there are only modest deviations for the longer period 
1880/1990 to 2025/2023 indicating that the parameters of the data change only 
adiabatically over the years such that it is plausible to assume 
thermalization for a given year. 

In Fig.~\ref{fig4}, we compare specifically the cases of female and 
male names of US 2025 and FR 2023. For both, the 
Lorenz and Pareto curves for the female and male cases are 
indeed very close. Appendix Fig.~\ref{figA4}, shows the same comparison 
for US 1880 and FR 1900 with a good match for FR 1900 but with 
some deviations for US 1880. 

\section{Results for the RJE model}
\label{sec4}

In this section, we compare results from the RJE model with real 
Lorenz and Pareto curves for certain cases of name frequencies 
(see also \cite{wth2} with more details on the RJE model and 
analytical results for both type of curves). 
For this, we compute an optimal value of the parameter $a$ such 
that the geometric curve distance 
between Lorenz RJE curve and data curve is minimized 
under the constraint that for each value of $a$ the other parameter $\eps$ 
is determined such that the Gini coefficient $G$ is identical for both cases. 
We have applied this procedure to many of the data sets discussed in 
the last section and the matching of the resulting RJE curves with real 
data curves is typically very good with an average geometrical curve 
distance below $5\times 10^{-3}$ and sometimes with values 
$\sim 2\times 10^{-3}$. 
As concrete examples, we show the comparison of RJE and real data curves 
for the cases of US F 2025 and FR F 2023 in Fig.~\ref{fig5} with 
parameter values of $a$ and $\eps$ given in the caption. 
Further examples are shown in Appendix Figs.~\ref{figA5} 
(US F 1880, FR F 1900), 
\ref{figA6} (US M 2025, FR M 2023) and \ref{figA7} (US M 1880, FR F 1900). 

For all  these  cases, the red data Lorenz curves are essentially identical 
on graphical precision with the blue RJE Lorenz curves. 
Furthermore, also the Pareto curves (for the same parameters 
$a$ and $\eps$ as for the Lorenz curve) agree very well with some deviations for 
a few largest values of $w_m$. We remind that the Pareto curves 
are shown in a double logarithmic representation that highlights different 
regions of $w_m$, i.e. the rich phase of largest $w_m$ while the Lorenz 
curves, used to determine the optimal $a$ values,  emphasize small and 
modest values of the cumulated ``wealth'' (normalized name frequency). 
However, in this context it is important to mention that while the 
Lorenz curves are scale independent (rescaling of $w_m$ units do not 
change them) the Pareto curves depend on a scale parameter and in 
all figures with Pareto curves, we show the ratio $w_m/\langle w_m\rangle$ 
where $\langle w_m\rangle$ is the average wealth (name frequency). 
For real data this quantity takes typical values in the interval 
$\sim$50-200 (see figures of last section) while for the RJE model, 
we have by construction $\langle w_m\rangle=\eps$ with $<1$ (typical 
values $\sim 10^{-3}$-$10^{-2}$). 
This is due to the fact that for the RJE spectrum we have chosen
for simplicity the global energy band width $B=E_{\max}=w_{\max}=1$ while 
for real data we have typically $w_{\max}\sim 10^3$-$10^4$. 
We note that the good agreement of the Pareto curves is obtained directly 
without any fit for the scale parameter provided the appropriate values 
of $\langle w_m\rangle$ for each case are used. 

In Fig.~\ref{fig6}, we show a collection of the RJE Lorenz curves 
of female US names for the years 1880 to 2020 (with a step of 
10 years) as a density color plot. The values of the optimal 
obtained RJE parameters are in the intervals $3<a<8.6$ and 
$2.9\times 10^{-3} < \eps < 2.4\times 10^{-2}$. 
Globally, one recognizes a rather 
strong ``poor'' (blue) phase (with RJ condensation) for all years (note the 
non-linear color scale to enhance small values). However, for the 
years between 1920 to 1960 the poor/blue zone is even a bit larger 
which can also be seen at the two lowest pink and blue Lorenz 
curves in the bottom left panel of Fig.~\ref{fig2} with maximal 
Gini coefficients $G\approx 0.94$. For example for 1940 50\% of names with 
lowest frequency were used for only 0.5\% of female new born in the US 
while the names with top 10\% (1\%) of frequencies were uses 
for 93\% (53\%) (these values for 1940 are even more extreme than those 
for US F for 2025 given in the introduction).

Appendix Fig.~\ref{figA8}, shows a similar density color plot for female 
FR names for 1900 to 2020 (step of 10 years) 
with $3<a<6.2$ and $2.7\times 10^{-3}<\eps<1.3\times 10^{-2}$. 
Here, we see that 
the poor/blue zone is maximal for 1950 but a more precise look reveals 
that for 1950 the rich/red zone zone is a bit larger in comparison to 
1960 and 1970 which have actually slightly larger values of the Gini 
coefficients than the 1950 FR F Lorenz curve.

Fig.~\ref{fig7}, shows the dependence of the Gini coefficient 
on the year for all cases US F/M (1880 to 2025) and FR F/M (1900 to 2023). 
For the US and FR female data the curves of Fig.~\ref{fig7} approximately confirm 
the above observations of Figs. \ref{fig6} and \ref{figA8}.
Globally, the Gini coefficients is maximal for the period between 
1920 and 1980 with typical values $G\approx 0.93$-0.95 
while for the initial (1880-1910) and final years (2010-2025), we have 
$G\approx 0.85-0.9$. However, the global interval $[0.85,0.95]$ 
of $G$ variations is quite modest. 

\begin{figure}[h]
\begin{center}
\includegraphics[width=0.95\columnwidth]{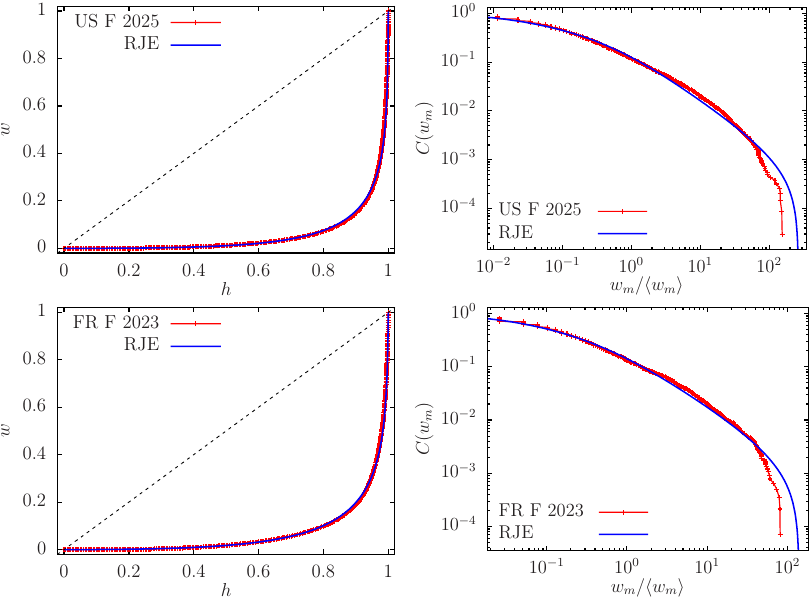}%
\end{center}
\caption{\label{fig5}
Lorenz curves (left panels) and 
Pareto curves (right panels) for the female name frequency 
of US 2025 (top) and FR 2023 (bottom) in the same style as in Fig.~\ref{fig2}. 
Red data points correspond to the data of \cite{govus,govfr} 
and blue curves correspond to the 
theoretical RJE curves obtained by matching Gini coefficients to determine 
$\eps$ and an optimal Lorenz curve fit to determine the parameter $a$. 
The optimal RJE parameters are 
$a=5.72,\ \eps=\langle w_m\rangle_{\rm RJE}=0.00376$, 
$\langle w_m\rangle_{\rm data}=87.9$  (US F 2025) 
and $a=5.18,\ \eps=\langle w_m\rangle_{\rm RJE}=0.0071$, 
$\langle w_m\rangle_{\rm data}=39.1$  (FR F 2023). 
See also Appendix Figs.~\ref{figA5},~\ref{figA6}.~\ref{figA7} for other cases of 
US F 1880, FR F 1900 (Fig.~\ref{figA5}),
US M 2025, FR M 2023 (Fig.~\ref{figA6}) and 
US M 1880, FR M 1900 (Fig.~\ref{figA7}). }
\end{figure}

\begin{figure}[h]
\begin{center}
\includegraphics[width=0.95\columnwidth]{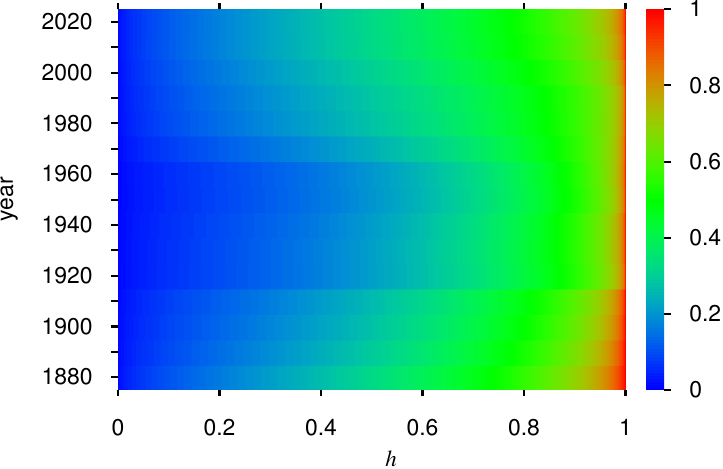}%
\end{center}
\caption{\label{fig6}
Density color plot of Lorenz curves for US F for the years 
1880 to 2020 (in steps of 10 years) 
obtained from the RJE model with optimal values of $a$ and $\eps$. 
The $x$-axis corresponds to the 
cumulated fraction of households/names ($h$) and the $y$-axis to
year. 
For a better visibility of small wealth values 
the numbers of the colorbar correspond 
to the quantity $w^{1/4}$ where $w$ is the cumulated fraction of 
wealth/name frequency. 
For all cases, the average curve distance between the real data and 
optimal RJE Lorenz curves is below $5\times 10^{-3}$. This applies also to 
Appendix Fig.~\ref{figA8}, where a similar density plot for FR F and the years 
1900 to 2020 is shown.}
\end{figure}

\begin{figure}[h]
\begin{center}
\includegraphics[width=0.95\columnwidth]{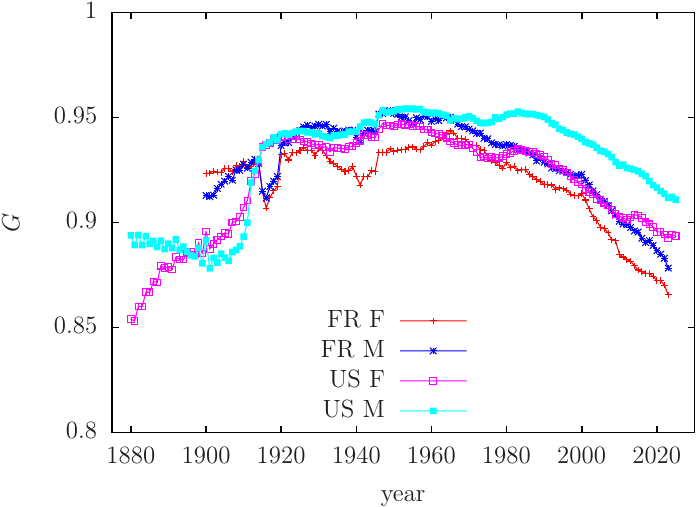}%
\end{center}
\caption{\label{fig7}
Year dependence of the Gini coefficient of the name Lorenz curves 
for the four cases of FR and US, both for female (F) and male (M) names.}
\end{figure}

\section{Time correlations}
\label{sec5}

\begin{figure}[h]
\begin{center}
\includegraphics[width=0.95\columnwidth]{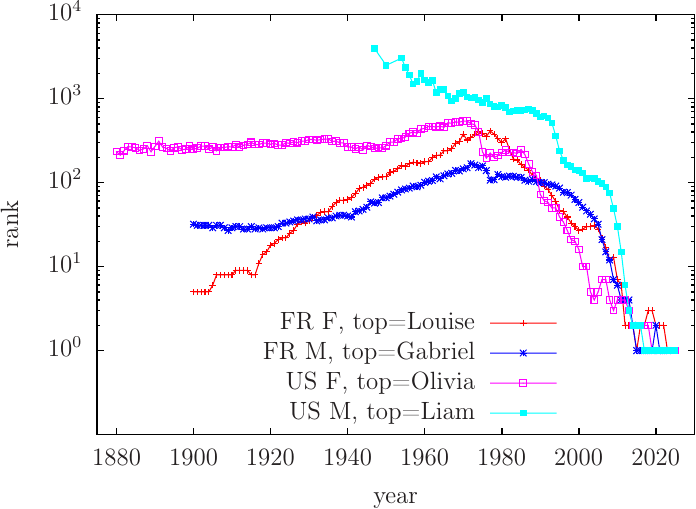}%
\end{center}
\caption{\label{fig8}
Year dependence of the rank of the top name of 2025 (2023) for US (FR) 
for the four cases of FR and US, both for female (F) and male (M) names. 
The obtained top names are given in the figure and the absence of data 
points for US M for years before 1947 indicates absence of the top name (Liam) 
in the corresponding year lists. }
\end{figure}

Globally, the very good agreement of both RJE Lorenz and Pareto curves with 
real data curves, deacribed in the previous section, confirms well the hypothesis 
of RJ thermalization (\ref{eqrj}) for the name frequency distribution 
(as it was the case for the wealth distributions $w_m$ \cite{wth1,wth2}).
However, this does not mean that there are no correlations between 
frequencies of names at different years. 
For each year the data sets of \cite{govus,govfr}
provide a ranking of the available names by their frequency value $f_m$. 
In Fig.~\ref{fig8}, we show for example 
the time dependence of the rank position $K$ of the two top US names 
Olivia/Liam at final year 2025 (for F/M) 
and also of the two top FR names Louise/Gabriel at 2023 (for F/M).
For a few years very close to the final year 
the four rank positions remain $K=1$ and for an interval of roughly 15
years around 2010\,-2025 they are still bounded by $K\le 3$. 
However, for earlier years they quickly increase to values 
$K\sim 10^2$-$10^3$ until 1970\,-1980. 
For US M the rank position increases further 
until 1947 to a value $K\sim 4\times 10^3$ 
and before 1947 the 2025 top name Liam even disappears from the list. 
For US F the rank position of Olivia saturates at earlier years at 
$K\sim 200$. 
For FR F, there is very long time revival, 
i.e. the name Louise was ranked at $K=5$ 
in 1900, it became almost forgotten with 
$K \approx 500$ in 1970 but it gained the top the position $K=1$ in 2023 
($K\le 3$ for 2015\,-2023). 
A bit less spectacular evolution took place
for the 2023 FR M top name Gabriel with 
the initial value $K\approx 30$ in 1900\,-1925, an increase to 
$K=168$ in 1972 and gaining the top position $K=1$ in 2015\,-2023 
(with the exception $K=2$ for 2020). 

Of course, such an evolution
does not contradict the thermodynamic distribution.
Indeed, in a gas of colliding atoms a certain atom
can gain a high kinetic energy and then loose it after a certain time due to 
several collisions corresponding to natural thermodynamic fluctuations.

To analyze further the time correlations of the name frequency data 
sets, we use the Pearson correlation coefficient \cite{pearson} 
which is defined for two given data sets $X_i$ and $Y_i$ of size $n_s$ as
\begin{equation}
\rho_{XY}=
\frac{\langle (X-\langle X\rangle)(Y-\langle Y\rangle)\rangle}
{\sigma_X\sigma_Y}=
\frac{\langle XY\rangle-\langle X\rangle\langle Y\rangle}
{\sigma_X\sigma_Y}
\label{defcorr1}
\end{equation}
where $\langle f(X,Y)\rangle=\sum_i f(X_i,Y_i)/n_s$ is the average over 
an arbitrary function $f(X,Y)$ and 
$\sigma_X=\sqrt{\langle X^2\rangle-\langle X\rangle^2}$, 
$\sigma_Y=\sqrt{\langle Y^2\rangle-\langle Y\rangle^2}$
is the standard deviation of $X$ and $Y$ respectively. 
Mathematically, $\rho_{XY}$ can in theory take values between $-1$ and 
$1$. Values close to $1$ ($-1$) indicate strong linear (anti-)correlations 
of the form $Y_i\approx c X_i+b$ with some coefficients $c,b$ 
and sign $c>0$ ($c<0$) while 
values close to 0 indicate small or absent correlations.

In this work, we have computed the Pearson correlation coefficient 
for the name frequencies between different years 
using the top 40 ranked names from 1910. 
Then the quantities $X_i$ and $Y_i$ 
represent the unshifted frequencies $f_m$ for two different years of these 
40 names (the absence or presence of a 
shift obviously does not affect the Pearson coefficient). 
If a name of this list does not appear in the list of a certain 
year, we attribute to this name the value $f_m=0$ (while other $f_m$ values 
for names present in a year list are always $\ge 5$ for US or $\ge 3$ for FR).

\begin{figure}[h]
\begin{center}
\includegraphics[width=0.95\columnwidth]{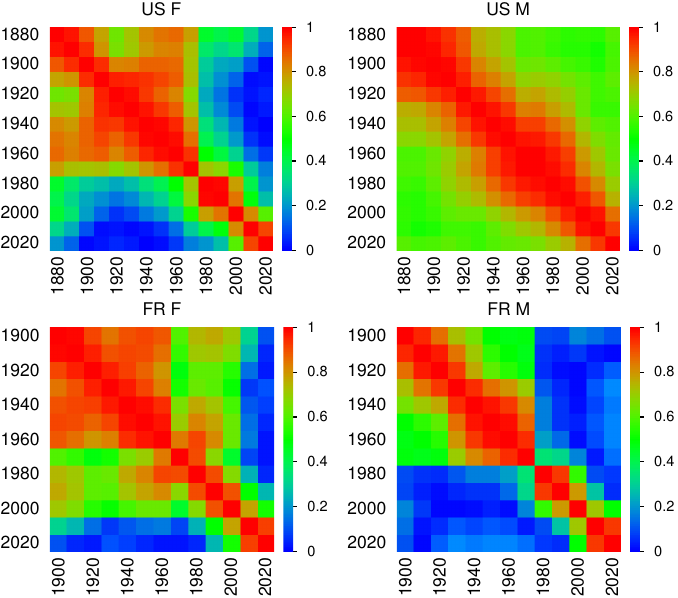}%
\end{center}
\caption{\label{fig9}
Color density plots of Pearson correlator (see text for definition) 
between different years for the name frequency 
for the four cases FR/US and F/M using top 40 names of 1910. 
The values of the colorbar indicate the correlator value. 
Since most correlator values are $\ge 0$ the colorbar is only 
shown for the interval $[0,1]$. A few cells with slightly negative 
values correspond mostly to blue color (or cyan, e.g. for two cells of FR M 
with strongest negative correlator value $-0.185$ between 1930 and 2020).}
\end{figure}

In Fig.~\ref{fig9}, we show this correlator as 
color density plots for the four usual cases (US/FR and F/M) and 
for the same sets of years $t$ used in Figs. \ref{fig6} and \ref{figA8},
i.e. from 1880 to 2020 for US and from 1900 to 2020 for FR 
with steps of 10 years. In Fig.~\ref{fig9}, most correlator values 
are positive and  only a few values are slightly negative (the strongest 
negative value is $-0.185$ for FR M). 

For US M, we observe globally quite strong correlations with 
values $\rho(\tau)=\rho_{f_m(t),f_m(t+\tau)}\approx 1$ 
for $\tau<20$-40 and minimal $\rho(\tau)\approx 0.5$ for 
$\tau\ge 80$. For the other three cases, we have 
dominant correlations inside the block $t\le 1960$-1970 
and decaying correlations for $t>1970$ on a scale of $\tau\approx 20$. 
In particular, correlations between $t<1990$ and $t>2000$ (or $t>1970$ for FR 
M) are very small with $\rho(\tau)\approx 0$. 
Concerning the block $t\le 1960$-1970, one can 
mention the examples of Marie and Jean who remain at the top position $K=1$ in
the years 1900-1950 for FR F/M. 
Also Mary is at the top position $K=1$ for the years 1880\,-1940 (US F)
and similarly for John for 1880\,-1920 (US M). 
We attribute this stability of top names in FR and US
to the christian religion influence
which drops after World War II. 

We remind, that the Pearson correlator coefficient is rather simple 
and that it is directly computed from the frequency values $f_m$ (and not the 
ranking index). If among the list of 40 names some of them 
drop out of the list for certain years, they get the frequency 
value $f_m=0$ which may artificially enhance the correlations between these 
years. Without going into details, we mention that we have also 
computed 
other correlator quantities based on local ranking (rank inside 
the group of 40 names) and global ranking (ranks of the 40 names in 
the full list for each year) or on the relative number of pairs with 
same sorting order between two years (Kendall correlator). 
These other correlator quantities essentially confirm 
the above observations for the Pearson correlator (for frequencies) 
even though they may provide a more complicated sub structure inside the block 
of $t\le 1960$. 

\section{Discussion} 
\label{sec6}

In this work, we analyzed the statistical distributions of given names 
available for more than hundred years 
at the government data bases of US and FR \cite{govus,govfr}.
We show that it is appropriate to characterize these 
distributions by the associated Lorenz and Pareto curves
broadly used in the context of wealth inequality. 
These curves 
remain remarkably stable on a scale of over 100 years
both for US and FR. Thus the Gini coefficient of Lorenz curves
varies only in the narrow interval 
$0.85 \leq G \leq 0.95$. The frequencies/occurrences/popularity of names
have two main components
with a fraction of about 1.2\%/1.0\% of cumulated frequencies for
50\% of less popular names and 85\%/88\% fraction of all frequencies
for top 10\% of names in US (e.g. for a typical case of 
female/male names of the year 2025).
For France in 2023, we have respectively 1.8\%/1.5\% (of cumulated 
frequencies) for the bottom 50\% fraction (of names) 
and 81\%/83\% for the top 10\% fraction (F/M).
For the top 1\% fraction of names we have respectively
43\%/47\% occurences for US and 37\%/38\% for FR in these years (F/M). 
For certain examples, there are even more extreme values, e.g. for US F 1940 
(see discussion of Fig.~\ref{fig6}). 
These kind of distributions are actually 
very similar to those of the wealth inequality in the world
with analogous fractions of poor and oligarchic population \cite{piketty2} 
and also similar values of optimal RJE parameters \cite{wth2}. 
In \cite{wth1,wth2} this wealth inequality is
explained on the basis of RJ thermal condensation
well known for various physical systems
\cite{picozzi1,picozzi2,ourfiber,trizac,satya,marsili}.
In this work, we show for names that the RJ condensation 
also provides a very good description of Lorenz and Pareto curves
on a scale of 100 years. Therefore, it is reasonable 
to assume that the phenomenon of RJ condensation is 
also the driving mecanism for the inequality
in the frequency or popularity of the given name distributions studied here. 
In this context, the thermalization in these distributions 
appears due to interactions of various preferences of names 
between certain agents or people taking place 
via books, magazines, radio and other media. 

Of course, one can argue that the fact that
the RJE model with two parameters fits
the real Lorenz and Pareto curves is not a sufficient argument
in the full favor of RJ thermalization theory.
However, we demonstrated in \cite{wth1,wth2,wth3,energy,election} 
and in this work
that the condensate and oligarchic phases exist for distributions
in variety of systems.
These systems are: wealth inequality in countries, Gross Domestic Product (GDP)
of countries, market capitalization of companies at stock exchange,
world trade, bitcoin transactions (see \cite{wth1,wth2});
energy and carbon emission distributions over countries \cite{energy},
distribution of  votes between parties in EU elections \cite{election}
and distribution of names as it is shown in this work. 
We also observe, that the Lorenz and Pareto curves in these systems
remain stable on time scales from 40 to 100 years.
Thus there should be a universal physical phenomenon
that is behind the striking similarity and stability of
distributions in these very different systems.
We argue that the RJ thermalization and condensation of classical fields
is in fact such  a universal phenomenon.
Indeed, thermalization is universal and it is extremely stable.
We consider that these arguments give the fundamental confirmation
of the validity of the RJ thermalization and condensation theory
in the above systems as it is described 
in this work and in \cite{wth1,wth2,wth3,energy,election}.

\section{Appendix}
\label{secapp} 

\setcounter{figure}{0} \renewcommand{\thefigure}{A\arabic{figure}} 
\renewcommand{\theHfigure}{Appendix.\thefigure}

Here we present additional Figures related to the material discussed in the main part of this work.

\begin{figure}[H]
\begin{center}
\includegraphics[width=0.95\columnwidth]{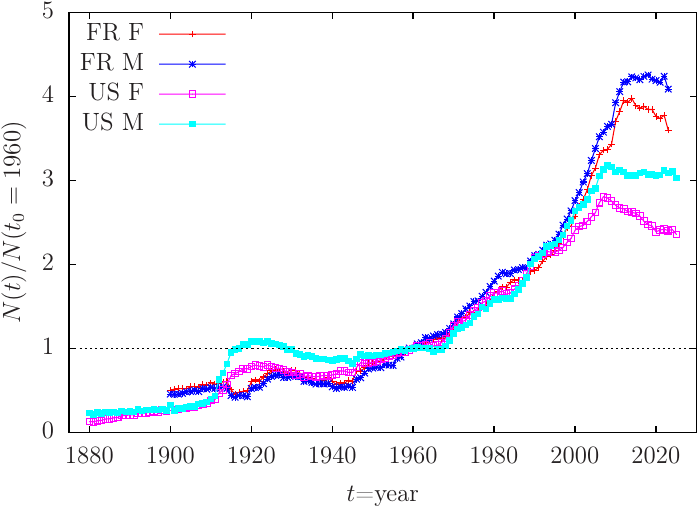}%
\end{center}
\caption{\label{figA1}
Year dependence of the number of names normalized to its value of 1960 
in the data sets of \cite{govus,govfr} 
for the four cases of FR and US, both for female (F) and male (M) names.}
\end{figure}

Appendix Fig.~\ref{figA1} shows the ratio $N(t)/N(1960)$ 
for the four cases US F/M and FR F/M 
where $N(t)$ is the number of names for the year $t$. 
The four curves collapse essentially to one curve for the interval 
$[1900,2005]$ with some deviations for US M between 1910 and 1940. 
The data for $N(t)$ can be roughly fitted by the exponential functions 
$N(t)=2270\times e^{0.0185(t-1960)}$ (FR F), 
$N(t)=1812\times e^{0.0206(t-1960)}$ (FR M), 
$N(t)=7477\times e^{0.0197(t-1960)}$ (US F) and 
$N(t)=5305\times e^{0.0183(t-1960)}$ (US M). 

\begin{figure}[H]
\begin{center}
\includegraphics[width=0.95\columnwidth]{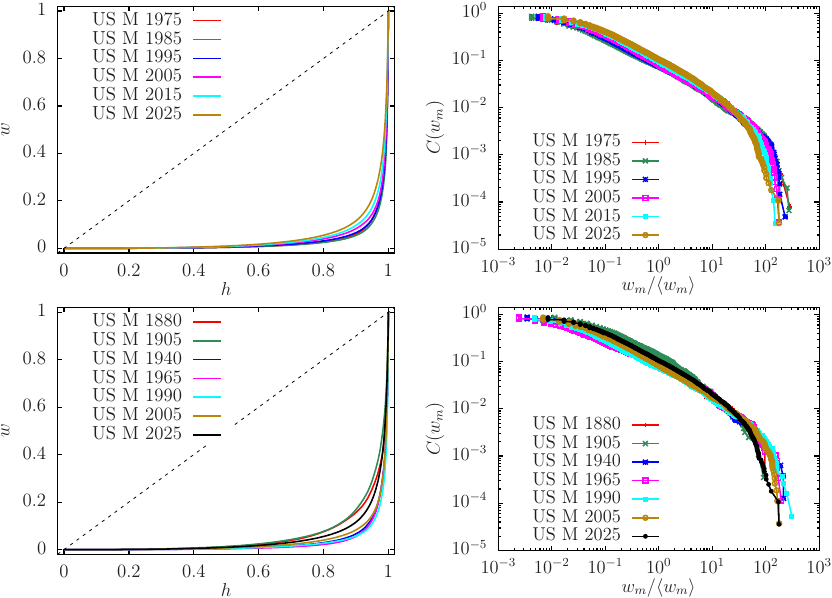}%
\end{center}
\caption{\label{figA2}
As Fig.~\ref{fig2} for US male names Lorenz and Pareto curves.
The Gini coefficients 
for the years 1975 to 2025 (1880 to 2025) are 
$G=0.948,0.952,0.944,0.934,0.924,0.911$
($G=0.894,0.883,0.943,0.949,0.95,0.934,0.911$)
and the average name frequency values are 
$\langle w_m\rangle = 241,239,179,144,131,118$
($\langle w_m\rangle = 99.4,88.1,289,415,211,144,118$).
}
\end{figure}

\begin{figure}[H]
\begin{center}
\includegraphics[width=0.95\columnwidth]{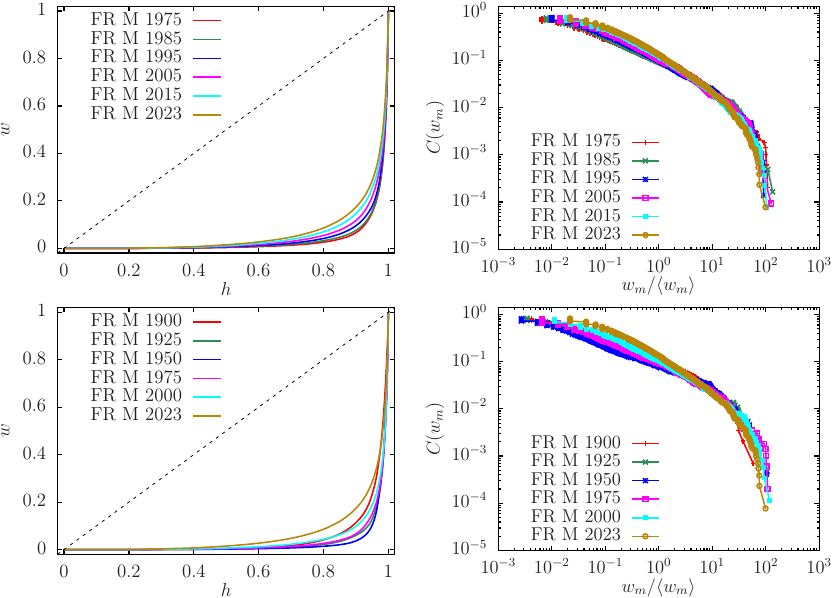}%
\end{center}
\caption{\label{figA3}
As Fig.~\ref{fig3} for FR male names Lorenz and Pareto curves.
The Gini coefficients 
for the years 1975 to 2023 (1900 to 2022) are 
$G=0.94,0.934,0.924,0.911,0.895,0.878$
($G=0.913,0.944,0.952,0.94,0.923,0.878$)
and the average name frequency values are 
$\langle w_m\rangle = 154,126,99.9,69.9,53.5,45.5$
($\langle w_m\rangle = 241,291,369,154,88.1,45.5$).}
\end{figure}

Appendix Figs.~\ref{figA2} and \ref{figA3}, show Lorenz and Pareto curves 
in the same style as Figs.~\ref{fig2} and \ref{fig3} for the cases of 
of US M (Fig.~\ref{figA2}) and FR M (Fig.~\ref{figA3}). The covered 
interval of years are the same as in Fig.~\ref{fig2} or in Fig.~\ref{fig3} 
respectively.

\begin{figure}[H]
\begin{center}
\includegraphics[width=0.95\columnwidth]{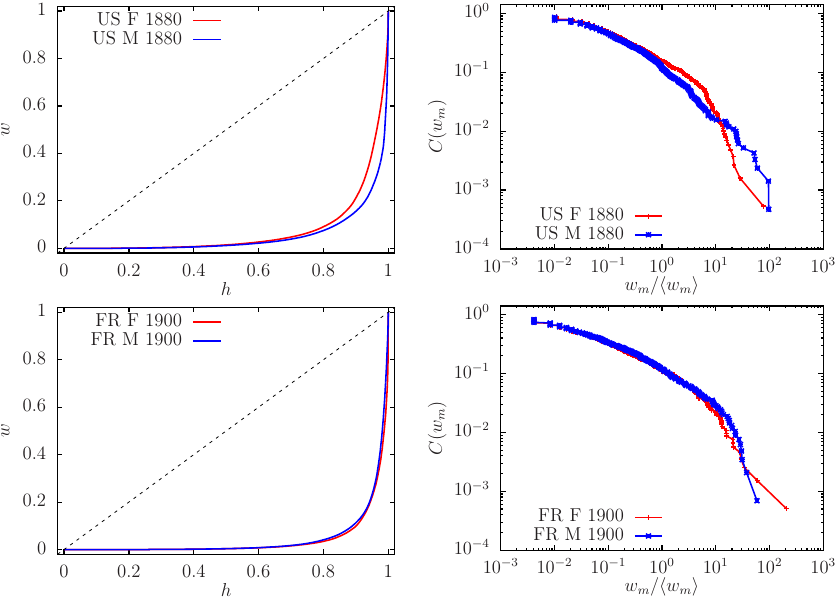}%
\end{center}
\caption{\label{figA4}
Comparison of Lorenz curves (left panels) and Pareto curves (right panels) 
between female names (red curves) and male names (blue curves). 
Top (bottom) panels correspond to data of US 1880 (FR 1900). }
\end{figure}

Appendix Fig.~\ref{figA4} provides, in the same style as 
Fig.~\ref{fig4}, a specific comparison between the 
cases of female and male names for US 1880 and FR 1900. While Lorenz 
and Pareto curves between both cases are very close for FR 1900 there 
are some visible deviations for US 1880. (Fig.~\ref{fig4} covers 
the cases of US 2025 and FR 2023 with close curves for both cases.)

\begin{figure}[H]
\begin{center}
\includegraphics[width=0.95\columnwidth]{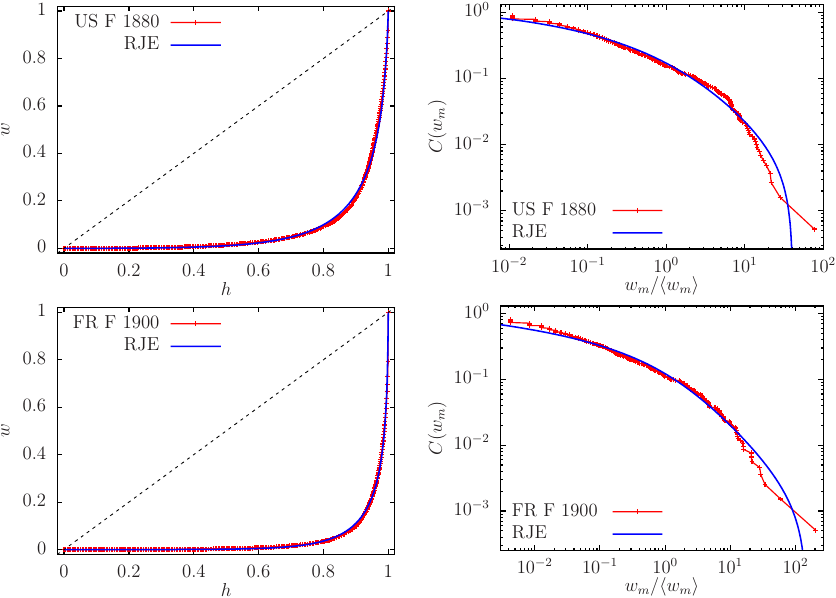}%
\end{center}
\caption{\label{figA5}
As Fig.~\ref{fig5} for the female name frequency 
of US 1880 (top) and FR 1990 (bottom). 
The optimal RJE parameters are 
$a=3.01,\ \eps=\langle w_m\rangle_{\rm RJE}=0.0244$, 
$\langle w_m\rangle_{\rm data}=91.6$  (US F 1880) 
and $a=4.17,\ \eps=\langle w_m\rangle_{\rm RJE}=0.00672$, 
$\langle w_m\rangle_{\rm data}=237$  (FR F 1900).}
\end{figure}

\begin{figure}[H]
\begin{center}
\includegraphics[width=0.95\columnwidth]{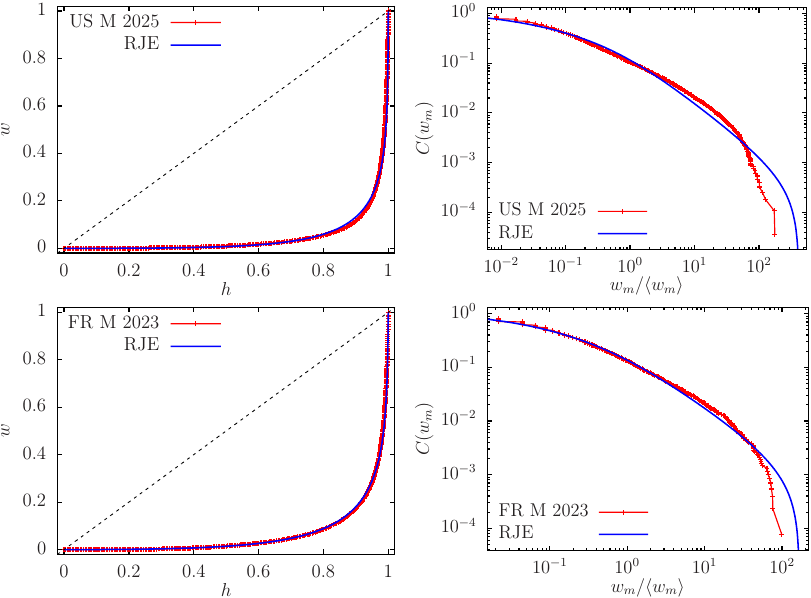}%
\end{center}
\caption{\label{figA6}
As Fig.~\ref{fig5} for the male name frequency 
of US 2025 (top) and FR 2023 (bottom). 
The optimal RJE parameters are 
$a=6.09,\ \eps=\langle w_m\rangle_{\rm RJE}=0.00238$, 
$\langle w_m\rangle_{\rm data}=118$  (US M 2025) 
and $a=5.26,\ \eps=\langle w_m\rangle_{\rm RJE}=0.00592$, 
$\langle w_m\rangle_{\rm data}=45.5$  (FR M 2023).}
\end{figure}

\begin{figure}[H]
\begin{center}
\includegraphics[width=0.95\columnwidth]{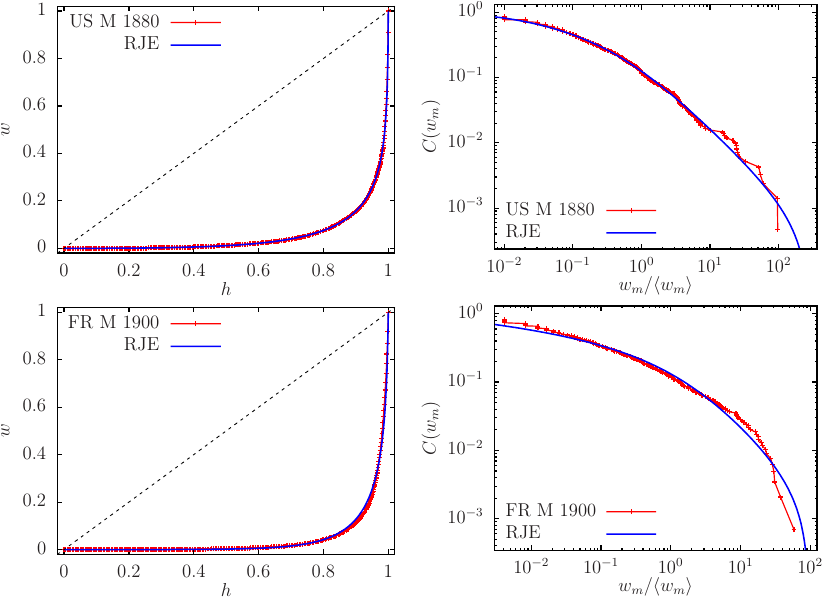}%
\end{center}
\caption{\label{figA7}
As Fig.~\ref{fig5} for the male name frequency 
of US 1880 (top) and FR 1990 (bottom). 
The optimal RJE parameters are 
$a=5.8,\ \eps=\langle w_m\rangle_{\rm RJE}=0.00358$, 
$\langle w_m\rangle_{\rm data}=99.4$  (US M 1880) 
and $a=3.62,\ \eps=\langle w_m\rangle_{\rm RJE}=0.0104$, 
$\langle w_m\rangle_{\rm data}=241$  (FR M 1900).
}
\end{figure}

Appendix Figs.~\ref{figA5}-\ref{figA7} compare RJE and real data 
(Lorenz and Pareto) curves in the same style as Fig.~\ref{fig5} 
for additional cases of 
US F 1880, FR F 1900 (Fig.~\ref{figA5},
US M 2025, FR M 2023 (Fig.~\ref{figA6} 
and US M 1880, FR F 1900) (Fig.~\ref{figA7}. 
For all these curves, one can observe a very good agreement 
between the real data and the RJE theory using optimal values 
of $a$ and $\eps$ given in the figure captions. 

\begin{figure}[H]
\vspace{1cm}
\begin{center}
\includegraphics[width=0.95\columnwidth]{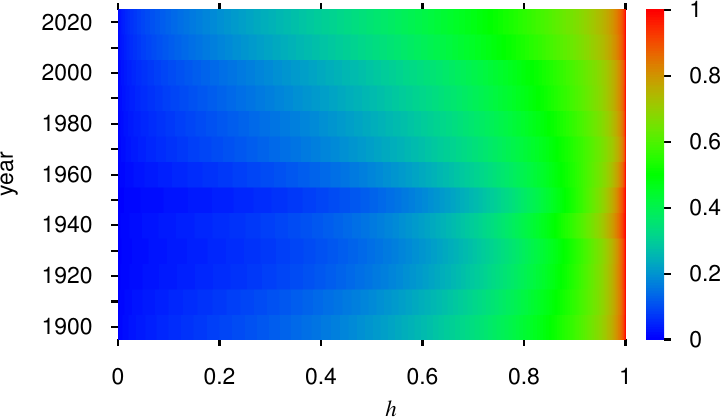}%
\end{center}
\caption{\label{figA8}
As Fig.~\ref{fig6} but for the Lorenz curves for FR F for the years 
1900 to 2020 (in steps of 10 years) 
obtained from the RJE model with optimal values of $a$ and $\eps$. }
\end{figure}

Appendix Fig.~\ref{figA8} shows a collection of the RJE Lorenz curves 
of female FR names for the years 1900 to 2020 (with a step of 
10 years) as a density color plot in the same style as Fig.~\ref{fig6} 
with optimal RJE parameters in the range 
$3<a<6.2$ and $2.7\times 10^{-3}<\eps<1.3\times 10^{-2}$. 

\vspace{1cm}

\begin{acknowledgments}
This research has been partially supported through the grant
NANOX $N^\circ$ ANR-17-EURE-0009 (project MTDINA) in the frame 
of the {\it Programme des Investissements d'Avenir, France}. 
This work was granted access to the HPC resources of
CALMIP (Toulouse) under the allocation 2026-P0110. 
\end{acknowledgments}

{\bf The data that supports the findings of this study are available within the article.
The used  data sets are obtained from Refs. \cite{govus,govfr}. }


\end{document}